\documentclass[prl,twocolumn,showpacs,letter]{revtex4}

\usepackage{graphicx}
\usepackage{amsmath}

\begin{document}

\title{Fermionic atoms in a Three Dimensional optical lattice: \\ Observing Fermi Surfaces,
Dynamics and Interactions}

\author{Michael K{\"o}hl$^\dag$, Henning Moritz, Thilo St{\"o}ferle, Kenneth G{\"u}nter, and
Tilman Esslinger}

\affiliation{Institute of Quantum Electronics, ETH Z\"{u}rich,
H\"{o}nggerberg, CH--8093 Z\"{u}rich, Switzerland}

\date{\today}

\begin{abstract}

We have studied interacting and non-interacting quantum degenerate
Fermi gases in a three-dimensional optical lattice. We directly
image the Fermi surface of the atoms in the lattice by turning off
the optical lattice adiabatically. Due to the confining potential
gradual filling of the lattice transforms the system from a normal
state into a band insulator. The dynamics of the transition from a
band insulator to a normal state is studied and the time scale is
measured to be an order of magnitude larger than the tunneling
time in the lattice. Using a Feshbach resonance we increase the
interaction between atoms in two different spin states and
dynamically induce a coupling between the lowest energy bands. We
observe a shift of this coupling with respect to the Feshbach
resonance in free space which is anticipated for strongly confined
atoms.

\end{abstract}


\maketitle

The exploration of quantum degenerate gases of fermionic atoms is
driven by the ambition to get deeper insight into long-standing
problems of quantum many-body physics, such as high temperature
superconductivity. Very recently, the cross-over regime between a
strongly interacting two-component Fermi gas and a molecular
Bose-Einstein condensate has been studied in harmonic traps
\cite{Regal2004,Bartenstein2004,Zwierlein2004,Kinast2004,Bourdel2004}.
These experiments mark a milestone towards the understanding of
superfluidity of fermionic atoms. However, the analogy to an
electron gas in a solid is limited since there the electrons
experience a periodic lattice potential. The lattice structure is
in fact a key ingredient for most models describing quantum
many-body phenomena in materials. It has been suggested that
strongly interacting fermionic atoms in optical lattices could be
employed for studies of high-Tc-superconductivity
\cite{Hofstetter2002}, Mott-insulating phases \cite{Rigol2003},
Bose condensation of fermionic particle-hole pairs \cite{Lee2004},
or interacting spin systems \cite{Santos2004}.

Here we report on an experiment bridging the gap between current
ultracold atom systems and fundamental concepts in condensed
matter physics. A quantum degenerate Fermi gas of atoms is
prepared in the crystal structure of a three dimensional optical
lattice potential created by three crossed standing laser waves.
The unique control over all relevant parameters in this system
allows us to carry out experiments which are not feasible with
solid-state systems.

It was conceived by D. Jaksch et al. that ultracold atoms exposed
to the periodic potential of an optical lattice are an almost
ideal realization of a Hubbard model \cite{Jaksch1998}. This model
is elementary to describe the quantum physics of many electrons in
a solid. It takes into account a single band of a static lattice
potential and assumes the interactions to be purely local
\cite{Hubbard1963}. Ultracold atoms in an optical lattice give a
very direct access to the underlying physics. The fundamental
parameters include the tunnel coupling between adjacent lattice
sites, the atom-atom interactions and the dimensionality of the
system. Previous experiments with far-detuned three dimensional
optical lattices \cite{DePue1999,Greiner2002a,Stoeferle2004} were
always carried out with bosonic atoms, and experiments with
fermions were restricted to a single standing wave
\cite{Modugno2003}. In the latter situation many atoms can reside
in each standing wave minimum but formation of a band insulator is
prevented by the weak transverse confinement. The observed
inhibition of transport\cite{Pezze2004} is due to localized states
and therefore differs qualitatively from the band insulator which
we create in the three dimensional optical lattice.

The experiments are performed in a modified apparatus previously
used to study bosonic rubidium atoms in optical lattices
\cite{Moritz2003,Stoeferle2004}. A mixture of bosonic $^{87}$Rb
and fermionic $^{40}$K atoms is captured in a magneto-optical
trap. For magnetic trapping we optically pump the potassium atoms
into the $|F=9/2, m_F=9/2\rangle$ and the rubidium atoms into the
$|F=2, m_F =2\rangle$ hyperfine ground state, with $F$ being the
total angular momentum and $m_F$ the magnetic quantum number. The
mixture is evaporatively cooled using microwave radiation to
selectively remove the most energetic rubidium atoms from the
trap. The potassium cloud is sympathetically cooled by thermal
contact with the rubidium atoms \cite{Modugno2002}. After reaching
quantum degeneracy for both species with typically $6\times 10^5$
potassium atoms at a temperature of $T/T_F=0.32$ ($T_F=260$\,nK is
the Fermi-temperature of the non-interacting gas) we remove all
the rubidium atoms from the trap. The potassium atoms are then
transferred from the magnetic trap into a crossed beam optical
dipole trap whose laser beams possess a wavelength of
$\lambda=826$\,nm and are focused at the position of the Fermi gas
to 1/$e^2$-radii of $50\,\mu$m (x-axis) and $70\,\mu$m (y-axis).
The initial trapping frequencies are $\omega_x=2 \pi \times
93$\,Hz, $\omega_y=2 \pi \times 154$\,Hz and $\omega_z=2 \pi
\times 157$\,Hz. When loading the optical trap we turn off the
magnetic confinement in such a way that a variable homogenous
magnetic field remains present. In the optical trap we prepare a
spin mixture with $(50 \pm 4)\%$ in each of the $|F=9/2,
m_F=-9/2\rangle$ and $|F=9/2, m_F=-7/2 \rangle$ spin states using
a sequence of two radio frequency pulses. By lowering the depth of
the optical trap on a time scale of 2 seconds we further
evaporatively cool the potassium gas. This is done at a bias
magnetic field of $B=227$\,G, which is well above the magnetic
Feshbach resonance centered at $B_0=202.1$\,G \cite{Regal2004} and
the s-wave scattering length between the two fermionic spin states
is $a=118 a_0$ ($a_0$ is the Bohr radius). At the end of the
evaporation we reach temperatures between $T/T_F=0.2$ and 0.25
with $5 \times 10^4$ to $2\times 10^5$ particles, respectively.

Prior to loading the atoms into the optical lattice we tune the
magnetic field to $B=(210 \pm 0.1)$\,G, such that the s-wave
scattering length between the two states vanishes. Then the
standing wave laser field along the vertical z-axis is turned on.
Subsequently, the optical dipole trap along the y-axis is turned
off and a standing wave laser field along the same axis is turned
on, followed by the same procedure along the x-axis. In order to
keep the loading of the atoms into the lattice as adiabatic as
possible the intensities of the lasers are slowly increased
(decreased) using exponential ramps with time constants of 10 ms
(25 ms) and durations of 20 ms (50 ms), respectively.

In its final configuration the optical lattice is formed by three
orthogonal standing waves with mutually orthogonal polarizations
and 1/$e^2$-radii of $50\,\mu$m (x-axis) and $70\,\mu$m (y-axis
and z-axis), which are derived from the same lasers as for the
optical dipole trap. The laser fields of the three beams have a
linewidth of the order of 10\,kHz and their frequencies are offset
with respect to each other by between 15 and 150\,MHz. The
resulting optical potential depth $V_{x,y,z}$ is proportional to
the laser intensity and is conveniently expressed in terms of the
recoil energy $E_r=\hbar^2 k^2/(2m)$, with $k=2 \pi / \lambda$ and
$m$ being the atomic mass. The lattice depth was calibrated by
modulating the laser intensity and studying the parametric
heating. The calibration error is estimated to be $<10\%$.

\begin{figure}[htbp]
  \includegraphics[width=\columnwidth,clip=true]{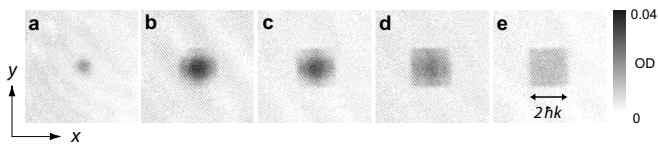}
  \caption{Observing the Fermi surface. Time of flight images
obtained after adiabatically ramping down the optical lattice. The
characteristic density increases from left to right. (a) 3500
atoms per spin state and a potential depth of the optical lattice
of $5\,E_r$. Images (b)-(e) were obtained with 15000 atoms per
spin state. The potential depths of the optical lattices were
$5\,E_r$ (b), $6\,E_r$ (c), $8\,E_r$ (d) and $12\,E_r$ (e). The
images show the optical density (OD) integrated along the
vertically oriented z-axis after 9\,ms of ballistic expansion.}
  \label{fig1}
\end{figure}

The potential created by the optical lattice results in a simple
cubic crystal structure and the gaussian intensity profiles of the
lattice beams give rise to an additional confining potential which
varies with the laser intensity. As a result, the sharp edges
characterizing the $T=0$ distribution function for the quasi
momentum in the homogeneous case \cite{Ashcroft1976} are expected
to be rounded off. A quantitative picture can be obtained by
considering a tight-binding Hamiltonian to describe
non-interacting fermions in an optical lattice with an additional
harmonic confinement \cite{Rigol2004}. At $T=0$ the inhomogeneous
system is characterized by the total atom number $N$ and by the
characteristic length $\zeta$ over which the potential shift due
to the harmonic confinement equals the tunnel coupling matrix
element $J$. One finds $\zeta_\alpha=\sqrt{2 J/m
\omega_\alpha^2}$, with the frequencies of the external harmonic
confinement given by $\omega_\alpha$ ($\alpha =x, y, z$). The
density distribution scaled by $\zeta_\alpha$ and the momentum
distribution of the atoms in the lattice only depend on the
characteristic density $\rho_c=\frac{N d^3}{\zeta_x \zeta_y
\zeta_z}$, where $d$ is the lattice spacing \cite{Rigol2003}. For
a three-dimensional lattice with $20\times 20\times20$ sites we
have numerically calculated the characteristic density for the
onset of a band insulator to be $\rho_c \simeq 60$. For this value
of $\rho_c$ the occupation number at the center of the trap is
larger than 0.99. It has been pointed out that a fermionic band
insulator in an optical lattice with confining potential
constitutes a high fidelity quantum register \cite{Viverit2004}.

\begin{figure}[htbp]
  \includegraphics[width=\columnwidth,clip=true]{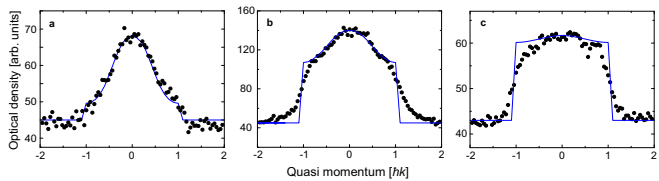}
  \caption{Analysis of the density distributions. The dots are cuts
through the measured density distribution for quasi momentum
$q_y=0$ after adiabatically ramping down the optical lattice. (a)
Normal state with $\rho_c=14.5$, (b) band insulator with
$\rho_c=137$, (c) band insulator with  $\rho_c=2500$. We have
numerically calculated the momentum distribution function of
fermions in the lowest band of a three-dimensional lattice with
$20\times 20\times 20$ sites and characteristic lengths
$\zeta_x/d=3.2$, $\zeta_y/d=2.6$,  $\zeta_z/d=2.5$ ((a) and (b))
and $\zeta_x/d=1$, $\zeta_y/d=0.8$, $\zeta_z/d=0.8$ (c), assuming
zero temperature (solid lines). Experimental data of (c) are
averaged over 5 images. Imperfect adiabaticity during the
switch-off of the optical lattice may cause the rounding-off of
the experimental data at the edge of the Brillouin zone in (b) and
(c). The calculated momentum distribution function is scaled to
match the experimental data using identical scale factors for all
graphs. }
  \label{fig2}
\end{figure}

In the experiment we probe the population within the first
Brillouin zones by ramping down the optical lattice slowly enough
for the atoms to stay adiabatically in the lowest band whilst
quasi-momentum is approximately conserved \cite{Greiner2001}. We
lower the lattice potential to zero over a timescale of 1\,ms.
After 1\,ms we switch off the homogeneous magnetic field and allow
for total of 9\,ms of ballistic expansion before we take an
absorption image of the expanded atom cloud. The momentum
distribution obtained from these time of flight images, shown in
Fig. \ref{fig1}, reproduces the quasi-momentum distributions of
the atoms inside the lattice. With increasing characteristic
density the initially circular shape of the Fermi surface develops
extensions pointing towards the Bragg planes and finally
transforms into a square shape completely filling the first
Brillouin zone deeply in the band insulator. We have observed
population of higher bands if more atoms are filled into the
lattice initially. In Fig. \ref{fig2} the experimental data for
momentum distributions along the line with quasi-momentum $q_y=0$
are compared to the results of numerical simulations using the
same characteristic densities.

When imaging the cloud along the x-direction we find a homogeneous
filling of the band in the vertical (z-) direction, probably due
to the change in the harmonic confinement while loading the
lattice combined with the presence of gravity. This asymmetry
between the horizontal x-, y-, and the vertical z-directions
vanishes when the gas approaches the band insulating regime. We
have examined the level of adiabaticity of our loading scheme into
the optical lattice by transferring the atoms from the band
insulator back into the crossed beam dipole trap. There we find a
temperature of 0.35\,$T_F$ when the initial temperature prior to
loading into the lattice was 0.2\,$T_F$.

\begin{figure}[htbp]
  \includegraphics[width=.7\columnwidth,clip=true]{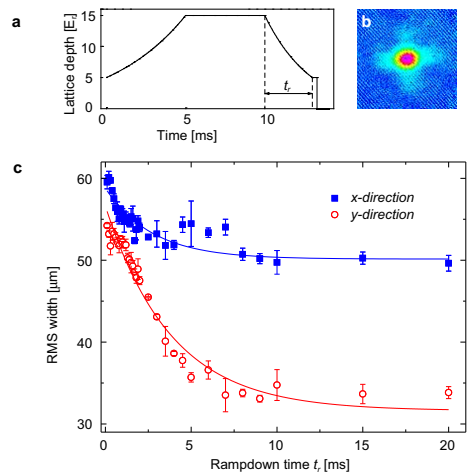}
  \caption{Restoring phase coherence. (a) Control sequence for the
depth of the optical lattice. (b) Pseudocolor image of the
momentum distribution after releasing the atoms from the initial
optical lattice of 5\,$E_r$ and 7\,ms ballistic expansion. It
reveals the central momentum peak and the matter wave interference
peaks at $\pm 2 \hbar k$. The data are averaged over 5 repetitive
measurements. (c) Width of the central momentum peak obtained from
Gaussian fits to the atomic density distribution. The initial
width is determined by the momentum spread of an atom localized in
the vibrational ground state of a lattice well. The $10\%$
difference in this size comes from slightly different
magnifications of the imaging system in the two orthogonal
directions. The difference in the asymptotic values of the width
can most likely be attributed to the loading sequence of the
lattice and to the asymmetry of the confining potentials due to
the different beam waists. The error bars show the statistical
error of 4 repetitive measurements.}
  \label{fig3}
\end{figure}

We have studied the dynamic response of the non-interacting Fermi
gas to a change in the characteristic density from a value deep in
the band insulating regime to a value below. In the latter regime
the fermions are delocalized over several sites of the optical
lattice and an interference pattern is observed when the atoms are
abruptly released from the lattice. The width of the interference
peaks is a measure of the length scale over which the atoms are
delocalized in the lattice or, equivalently, their coherence
length. We change the characteristic density in situ by varying
the strength of the lattice laser beams. Starting from an initial
characteristic density of $\rho_c=16$ in an optical lattice with a
potential depth of $5\,E_r$ we create a band insulator with a
characteristic density of  $\rho_c=2700$ at a potential depth of
$15\,E_r$. After holding the atoms for 5 ms we reduce the
potential depth back to $5\,E_r$, using an exponential ramp with
duration and time constant $t_r$. This is followed by a rapid
switch off of the lattice (see Fig. 3a). We measure the width of
the central momentum peak in the time of flight images for
different durations $t_r$ and obtain the time scales $\tau_x=(2.7
\pm 0.4)$\,ms and $\tau_y=(3.8 \pm 0.3)$\,ms in the x- and
y-direction, respectively. This corresponds to approximately ten
times the timescale for tunnelling given by $h/2zJ$ at a potential
depth of $5\,E_r$, where $z$ is the coordination number of the
lattice. This non-trivial dynamics appears to be significantly
slower than the time scale measured for the transition of a
Mott-insulating state to a superfluid state using bosonic atoms in
an optical lattice \cite{Greiner2002a}. The comparatively slow
dynamics of delocalization of the fermions when approaching the
normal state is most likely due to Pauli blocking which prevents
tunneling of atoms in regions where the lowest band is full and
the atoms are well localized.

\begin{figure}[htbp]
  \includegraphics[width=.7\columnwidth,clip=true]{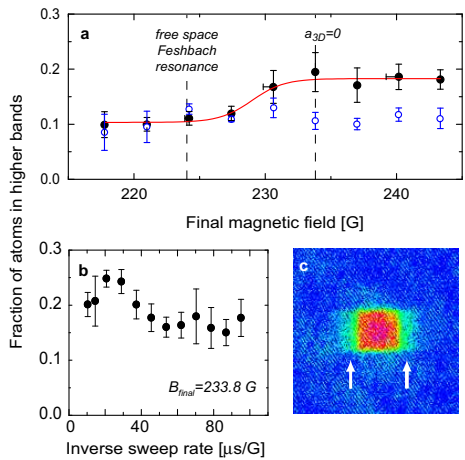}
  \caption{Interaction-induced transition between Bloch bands. (a)
Transferring fermions into higher bands using a sweep across the
Feshbach resonance (filled symbols). The inverse magnetic field
sweep rate is $12\,\mu$s/G. The line shows a sigmoidal fit to the
data. The open symbols show a repetition of the experiment with
the atoms prepared in the spin states $|F=9/2, m_F=-9/2\rangle$
and $|F=9/2, m_F=-7/2\rangle$ where the scattering length is not
sensitive to the magnetic field. The magnetic field is calibrated
by rf spectroscopy between Zeeman levels. Due to the rapid ramp
the field lags behind its asymptotic value and the horizontal
error bars represent this deviation. (b) Fraction of atoms in
higher bands for a final magnetic field of 233 G for different
magnetic field sweep rates. The vertical error bars show the
statistical error of 4 repetitive measurements. (c) Momentum
distribution for a final magnetic field of $B=233$\,G and a
$12\,\mu$s/G sweep rate. Arrows indicate the atoms in the higher
bands.}
  \label{fig4}
\end{figure}

We investigate the interacting regime in the lattice starting from
a non-interacting gas deep in a band insulator with $V_x=12\,E_r$
and $V_y=V_z=18\,E_r$ and corresponding trapping frequencies of
$\omega_x=2 \pi \times 50$\,kHz and $\omega_y=\omega_z=2 \pi
\times 62$\,kHz in the individual minima. A short radio-frequency
pulse is applied to transfer all atoms from the $|F=9/2,
m_F=-7/2\rangle$ into the $|F=9/2, m_F=-5/2\rangle$ state, with
the atoms in the $|F=9/2, m_F=-9/2\rangle$ remaining unaffected.
We ramp the magnetic field with an inverse sweep rate of
$12\,\mu$s/G to different final values around the Feshbach
resonance (see Fig. 4a) located at $B=224$\,G \cite{Regal2003}.
The sweep across the Feshbach resonance goes from the side of
repulsive interactions towards the side of attractive
interactions. When using this direction of the sweep there is no
adiabatic conversion to molecules. After turning off the optical
lattice adiabatically and switching off the magnetic field we
measure the momentum distribution. To see the effect of the
interactions we determine the fraction of atoms transferred into
higher bands. For final magnetic field values well above the
Feshbach resonance we observe a significant increase in the number
of atoms in higher bands along the weak axis of the lattice,
demonstrating an interaction-induced coupling between the lowest
bands. Since the s-wave interaction is changed on a time scale
short compared to the tunnelling time between adjacent potential
minima we may regard the band insulator as an array of harmonic
potential wells. It has been shown that increasing the s-wave
scattering length for two particles in a harmonic oscillator
shifts the energy of the two-particle state upwards until the next
oscillator level is reached \cite{Busch1998}. In our case this
leads to a population of higher energy bands. The fraction of
atoms transferred could be limited by the number of doubly
occupied lattice sites and tunnelling in the higher bands. The
number of doubly occupied sites could be measured by studying the
formation of molecules in the lattice. In addition, we observe a
shift of the position of the Feshbach resonance from its value in
free space to larger values of the magnetic field (see Fig.
\ref{fig4}a), which has been predicted for tightly confined atoms
in an optical lattice \cite{Fedichev2004}. This mechanism for a
confinement induced resonance is related to the phenomenon
predicted for one-dimensional quantum gases \cite{Olshanii1998}
which has as yet escaped experimental observation. For a
quantitative description of this strongly interacting Fermi gas on
a lattice a multi-band Hubbard model could be considered but these
are even in the static case notoriously difficult or even
impossible to solve with present methods \cite{Troyer2004}.

In conclusion we have created a fermionic many-particle quantum
system on a lattice. We have demonstrated the dynamic control over
the parameters of the system such as filling and interactions
which is not feasible in solid state systems. For the
non-interacting static regime we find good agreement between our
measurements and a theoretical model. Both the dynamic
measurements and the strongly interacting case pose challenges for
the present theoretical understanding of many-particle fermionic
systems on optical lattices.

We would like to thank G. Blatter, C. Bruder, H. P. B\"uchler, S.
Jonsell, A. Muramatsu, M. Rigol, C. Schori, P. T\"orm\"a and M.
Troyer for insightful discussions, and SNF, SEP Information
Sciences and QSIT for funding.

\end{document}